 \documentstyle[epsf,prl,aps,twocolumn,floats]{revtex}

\begin{document}

\flushbottom

\draft

 \twocolumn[\hsize\textwidth\columnwidth\hsize\csname @twocolumnfalse\endcsname

\title{
Two distinct metallic bands associated with monatomic Au wires on the
Si(557)-Au surface }

\author{Daniel S\'anchez-Portal$^{1,2}$\cite{byline}, Julian D. Gale$^3$,
Alberto Garc\'{\i}a$^4$ and, Richard M. Martin$^1$}

\address{
$^1$Department of Physics and Materials Research Laboratory,
 University of Illinois, Urbana, Illinois 61801\\
$^2$ Departamento de F\'{\i}sica de Materiales, 
Facultad de Qu\'{\i}mica, UPV/EHU, Apdo. 1072, E-20080 San Sebasti\'an, Spain \\
$^3$ Department of Chemistry, Imperial College of Science, Technology 
and Medicine, South Kensington SW7 2AY, London UK \\
$^4$ Departamento de F\'{\i}sica de la Materia Condensada, Universidad del 
Pa\'{\i}s Vasco, Apdo. 644, E-48080 Bilbao, Spain} 

\date{\today}

\maketitle

\begin{abstract}

The Si(557)-Au surface, containing monatomic Au wires parallel to the
steps, has been proposed as an experimental realization of an ideal
one-dimensional metal.  In fact, recent photoemission experiments on
this system (Nature {\bf 402}, 504 (1999)) found two peaks that were
interpreted in terms of the spin-charge separation in a Luttinger
liquid.  Our first-principles density functional calculations reveal
{\it two metallic bands} associated with Au-Si bonds, instead of the
single band expected from the Au 6$s$ states, providing an alternative
explanation for the experimental observations.

\end{abstract}



]

It has been recognized for several decades now that a one-dimensional
metal should considerably deviate from the Fermi liquid
behavior\cite{Tomonaga,Luttinger,Haldane,Fisher}: quasiparticles would
be replaced by two distinct collective excitations involving spin and
charge, and some of the system's properties would exhibit
characteristic power-law dependences.  Despite this theoretical
understanding, the experimental observation of the so-called Luttinger
liquid behavior has proven quite elusive, due in large part to the
lack of well-characterized one-dimensional systems.  Only very
recently, signatures of Luttinger liquid behavior have been observed
by means of transport measurements in carbon nanotubes\cite{NT}, in
edge states of the fractional quantum Hall effect\cite{Hall}, and in
quantum wires with a single conducting channel\cite{wires}, and by
optical conductivity measurements in organic conductors\cite{organic},
and in the quasi-one-dimensional CuO chains in
PrBa$_2$Cu$_3$O$_8$\cite{PrBaCuO}.

More direct evidence of the Tomonaga-Luttinger behavior can be
provided by angle-resolved photoemission experiments, in which the
spin-charge splitting of the low-energy excitations can, in principle,
be measured.  This has been tried for several systems containing
quasi-one-dimensional metallic chains\cite{SrCuO,LiMoO}.  However, in
most cases the results were inconclusive due to the lack of energy and
angular resolution\cite{Xue}, the difficulties to correctly align the
chains, and other uncertainties associated with the crossover of
dimensionality in the bulk samples.  To avoid some of these problems,
Segovia {\it et al.}\cite{NatureSegovia} studied the photoemission
from a monophase sample of the stepped Si(557)-Au surface, in which
the gold atoms are believed to form single rows running
parallel to the edge of each step. For this system, the measured
spectra were assigned to a single one-dimensional band derived from
the gold 6$s$ states.  The observed splitting of this band close to
the Fermi level was interpreted in terms of the spin-charge separation
in a one-dimensional electron liquid, providing the first direct
measurement of two distinct peaks for spinons and holons.  If such
behavior is intrinsic to Au wires, it could also apply to monatomic
gold wires fabricated using a scanning tunneling microscope (STM) and
to mechanically controllable break junctions\cite{Natureswires,PRL}.

In spite of this extraordinary observation, very little is still known
about the atomic and the electronic structure of the Si(557)-Au
surface. In this context, theoretical studies can be useful to
complement and guide experimental efforts. With this aim, we present
in this Letter a first-principles density-functional study of the
electronic structure of simple models of this stepped surface.  We
obtain the surprising result that {\it two metallic bands}, not just
one, are associated with the monatomic gold chains in this
system. This could explain the photoemission data of Segovia {\it et
al.}\cite{NatureSegovia} without the need to appeal to electron
correlation effects in the monatomic gold chains, and seems to be in
agreement with more recent experimental evidence\cite{privateHimpsel}.

Our calculations have been performed with the SIESTA
code\cite{SIESTA}, which performs standard pseudopotential
density-functional calculations using a linear combination of atomic
orbitals (LCAO) as a basis set.  We have used Troullier-Martins
pseudopotentials\cite{Troullier-Martins}, the local density
approximation for the exchange-correlation functional, and a
double-$\zeta$ polarized (DZP)\cite{SIESTA,basis} basis set for all
the atomic species. A Brillouin zone sampling of 10 inequivalent
k-points, and a real-space grid equivalent to a plane-wave cut-off of
100~Ry (up to 40 k-points and 200~Ry in the convergence tests) have
been employed. These parameters guarantee the convergence of the total
energy, for a given basis, within $\sim$0.02~eV/Au (due to the large
supercells this becomes $\sim$0.5~meV per atom).

The Si(557)-Au surface presents a (5$\times$1) reconstruction formed
after the deposition of $\sim$0.2 monolayers (ML) of gold on top of a
vicinal Si(111) surface, as described in Ref\cite{NatureSegovia}.  The
misorientation is chosen along the [$\overline{1}$~$\overline{1}$ 2]
direction and therefore the steps, with a width of five Si(111)
1$\times$1 unit cells, are parallel to the [1$\overline{1}$0]
direction.  A single row of gold atoms is supposed to run parallel to
each step, consistent with the gold coverage.  However, the detailed
structure of the Si(557)-Au surface is unknown. It is probably related
to the better studied (5$\times$2) reconstruction\cite{Mahony}, which
is formed on flatter substrates, and contains twice as much gold
($\sim$0.4 ML).  In the absence of precise structural information, we
have adopted in this study the simplest possible models consistent
with all the available experimental data.

\begin{figure}
\epsfxsize=\hsize\centerline{\epsffile{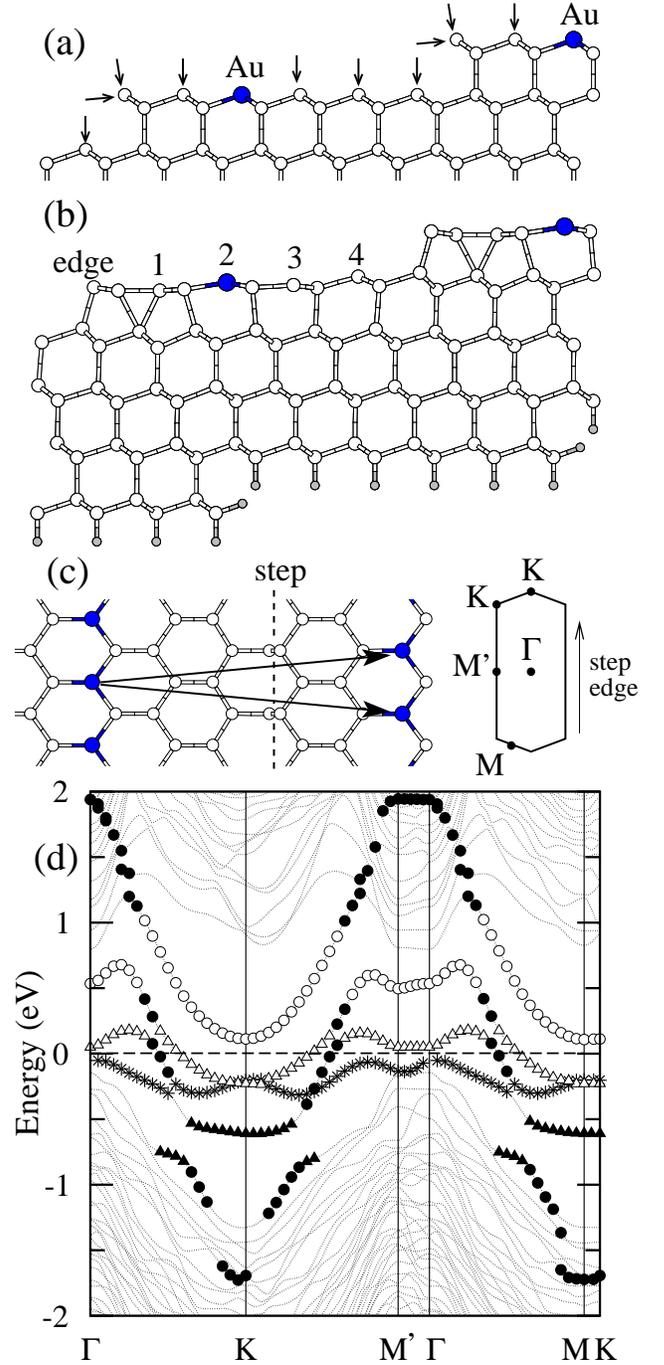} }
\caption[]{
(a) Unrelaxed atomic positions in a model for the Si(557)-Au surface
(hereafter model 2).  
(b) Same structure after relaxation.  (c) Top
view of the stepped surface, with the unit cell vectors indicated, and
the corresponding surface Brillouin zone.  (d) Electronic band
structure of the slab presented in panel (b). Energies are referred to
the Fermi level, and surface states have been marked with different
symbols according to their character (see text).}
\label{fig1}
\end{figure}

The initial and relaxed structures of one of these models, similar to
the one proposed in Ref.\onlinecite{NatureSegovia}, are shown in
Fig.\ref{fig1} (a) and (b), respectively.  In panel (a) the position
of the Si dangling bonds present initially have been schematically
indicated by arrows, while in panel (b) the rows of atoms are labelled
according to their distances to the step edge.  The typical slabs (see
Fig.\ref{fig1} (b)) used in our calculations contain 4 silicon
bilayers, with the Au atoms occupying some of the Si positions and
forming a single row parallel to the step edge.  Si atoms in the
bottom layer are saturated with hydrogen, and remain at the ideal bulk
positions during the relaxation process.  The lattice parameter
parallel to the surface is fixed at the calculated bulk value
(5.41~\AA ) to avoid artificial stresses.

During the relaxation, the step edge moves considerably from its
original position in order to saturate some of the surface
dangling-bonds with the creation of additional Si-Si bonds.  This also
causes some rebonding of the neighboring Si atoms.  The tendency of
the Si atoms at the surface to adopt a $sp^2$ hybridization, which
favors an expansion of the surface layer, also contributes to the
movement of the step.  These trends are common to the relaxed
structures of all the structural models considered here.

Table~\ref{table1} presents the total energy of several structures
relative to that of the configuration shown in Fig.\ref{fig1} (b).
The structures differ in the location of the Au chains, and are
labelled according to Fig.\ref{fig1} (b).  One can see that the
vicinity of the step edge (position 1) is a highly unfavorable
location for the gold row, whereas the energies associated to
positions 2, 3 and, 4 are almost degenerate within the precision of
our calculations.  (In addition, the band-structure features described
below are very similar for models 2,3, and 4.)  Consistent with these
results, recent STM experiments\cite{Shibata} indicate that position 2
could be the preferred location for the Au chain in narrow terraces.
Therefore, in the following we will concentrate on this structure
(model 2).

It has been proposed\cite{PRLHimpsel} that a Peierls dimerization of
the Au chains could lead to splittings of the bands.  Our calculations
show that in spite of the large Au-Au distance (3.83~\AA), the
dimerization is totally suppressed by the rigidity of Si structure.

The band structure near the Fermi level of the slab in model 2 is
shown in Fig.\ref{fig1}(d) with the lines defined by the
two-dimensional Brillouin zone in Fig.\ref{fig1} (c) (lines
$\Gamma$-K-M$^{\prime}$ are parallel and $\Gamma$-M almost 
parallel to the steps, while
$\Gamma$-M$^{\prime}$ is 
perpendicular to them).  The use of a
LCAO basis allows us to perform a Mulliken population
analysis\cite{Mulliken} to identify the main character of each band.
Though not totally unambiguous, this identification is specially clear
for surface states, which have a considerable weight only in the
outermost atoms of the slab.
In Fig\ref{fig1}(d) each surface state has been marked with different
symbols according to its origin: {\it i)} The rather
flat band marked with stars comes from the unsaturated bonds of the Si
atoms in the step edge.  This band, strongly localized at the edge,
is present, pinning the Fermi level, in all the models with Si
dangling bonds at the edge. It could be eliminated by
removing the extra Si atom at the edge and dimerizing the
remaining atoms in analogy to the Si(100)-2$\times$1 reconstruction,
with a doubling of the unit cell. However, our most stable relaxed
"edge-dimerized" model has a surface energy much higher
($\sim$0.4~eV/Au) than model 2.  {\it ii)} States marked
with open symbols correspond to dangling bonds on the
terraces. The partially occupied band marked with open triangles comes
from the unsaturated bonds in row 3, while the unoccupied band marked
with open circles is associated to those in row 4. {\it iii)} Two
"one-dimensional" bands, i.e. exhibiting an almost negligible
dispersion in the direction perpendicular to the Au chain, have very
similar amounts of gold character.  The one marked with solid
circles is more dispersive and crosses the Fermi energy, while the
other, indicated with solid triangles, is quite flat and fully
occupied.  The presence of these two bands contradicts the intuitive
picture that would assign a single band, coming from the 6$s$ state,
to the Au wire. In fact, our analysis reveals that a considerable
weight of these "gold bands" corresponds to the Si neighbors of the Au
atoms.

\begin{figure}
\epsfxsize=\hsize\centerline{\epsffile{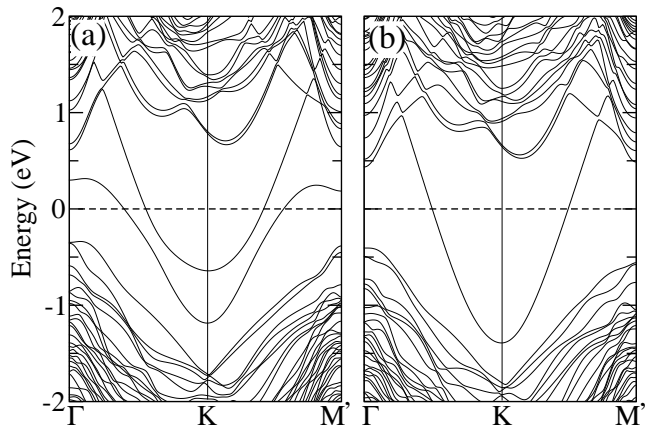}   }
\caption[]{Band structure of the relaxed slab: (a)~After the
saturation of all Si dangling bonds with hydrogen.  (b)~After one
additional hydrogen has been deposited on top of each gold atom.  }
\label{fig2}
\end{figure}

The most dispersive band in our calculations, with its bottom close to
the edge of the Brillouin zone, can be identified with the
experimentally observed one\cite{NatureSegovia}.  However, in this
system there is no report to date of the flat bands that we have
identified as due to unsaturated surface bonds (those marked with
stars and open triangles in Fig.\ref{fig1}(d)).  These bands are
probably eliminated in the real surfaces by surface reconstruction and
passivation. Even though we have no detailed information on the nature
of the reconstruction, we can account for its main effect, i.e., the
elimination of the dangling bonds, by saturating them with
hydrogen.\cite{footnoteRMM} This process is energetically very
favorable (we gain $\sim$1~eV per bond relative to the unsaturated
surface and the free H$_2$ molecule), and removes all the extra bands
from the gap region.  Fig.\ref{fig2}(a) shows the band structure of
the hydrogen-saturated surface, clearly revealing two metallic bands
associated to the Au wires, which correspond to those marked with
solid symbols in Fig.\ref{fig1}(d) for the unsaturated surface.
Interestingly enough, this band structure is in extremely good
agreement with a very recent photoemission experiment which seems to
favor a two-band scenario over the spin-charge
splitting\cite{privateHimpsel}.

We can also use the addition of hydrogen to make further predictions.
By further increasing the hydrogen partial pressure we can decorate
the Au wires with H, which is much less favorable than Si bond
saturation (the energy gain is reduced to $\sim$0.1~eV in this case).
The additional hydrogen atoms adopt a bridge position on top of the
Si-Au bonds, and the band structure (Fig.\ref{fig2} (b)) exhibits a
single metallic band.  Thus, by increasing the H content we have been
able to switch from a monatomic Au wire displaying two bands, to a
single band situation. This points to the conclusion that it is in
this hydrogen rich environment that signatures of Luttinger-liquid
behavior should be looked for.

\begin{figure}
\epsfxsize=\hsize\centerline{\epsffile{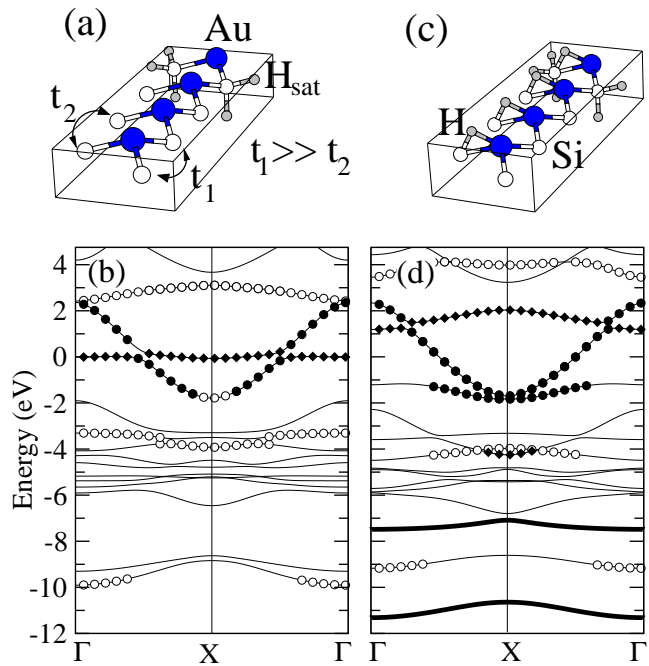}}
\caption[]{Simple models of (a) clean
and, (c) hydrogen-decorated, Au monatomic
wires in the Si(557)-Au surface.
Panels (b) and (d) present the energy bands of
the structures shown in (a) and (c), respectively.
Energies are referred to the Fermi level, and
states are marked according to
their atomic character (see text).}
\label{fig3}
\end{figure}

A simple model serves to explain the appearance of the two partially
filled bands while ignoring all the details of the surface
reconstruction.  In this model (Fig.\ref{fig3}(a)), we have only
considered the gold atoms and those Si atoms bound to them. The bonds
connecting with the rest of the surface have been saturated with
hydrogens H$_{sat}$ (only a few of them visible in the figure).  We
show the band structure of this simplified model wire in
Fig.\ref{fig3} (b). Open circles mark those states with main character
on the 6$s$ and 6$p$ orbitals of gold.  Clearly, the 6s states of Au
form a fully occupied band located ~3.6 eV below the Fermi level.
Closer to the Fermi energy we find two bands which, although
considerably hybridized with the gold orbitals, are basically derived
from the 3$sp$ "bonds" of the Si atoms. Therefore, while induced by
the presence of the Au wire, those are {\it silicon} bands.  The flat
band pinning the Fermi energy (solid diamonds) comes from the Si chain
schematically indicated in Fig.\ref{fig3}(a) by the hopping t$_2$. A
more dispersive band (solid circles) originates from the other Si
chain, where the interaction t$_1$ is stronger due to the more
favorable orientation of the "bonds", facing each other.  The
appearance of two metallic bands can be expected by a simple {\it
local counting of electrons} argument: There are four electrons in the
chain per Au atom, one electron is provided by the 6$s$ state of gold
and three come from the Si neighbors. Consequently, a metallic wire,
as observed in the experiments, in absence of considerable charge
transfer from other atoms in the surface, {\it must} display two
partially occupied bands.

The addition of hydrogen atoms decorating the chain
(Fig.\ref{fig3}(c)), produces the band structure presented in
Fig.\ref{fig3}(d), where the bands with high hydrogen character
are marked by thick solid lines.  In this case one electron is transferred
to the hydrogen, leaving unoccupied the flat Si band, and only the
most dispersive silicon band crosses now the Fermi level, in agreement
with our calculations for the H-saturated surfaces.

In conclusion, we have presented an {\it ab initio} study of total
energies and surface state bands of the Si(557)-Au surface. Our
results show that the Au chains lead to {\it two half-filled metallic
bands} associated with the Au-Si bonds that can be interpreted in a
simple model.  This is in accordance with very recent experimental
photoemission results\cite{privateHimpsel} and provide an alternative
to the spin-charge separation interpretation
proposed\cite{NatureSegovia} to explain the two-peak photoemission
spectra found for this system.
 
\acknowledgements

We thank A. Yazdani, E. Ortega and F. J. Himpsel for useful discussions
and NCSA for computational resources.
DSP and RMM. acknowledge support by Grants No. DOE 8371494,
and No. DEFG 02/96/ER 45439. DSP also acknowledge support from 
the Basque Government (Programa de Formaci\'on de Investigadores)
JDG would like to thank the Royal Society for a University Research
Fellowship and for funding. AG acknowledges support from Fundaci\'on
Ram\'on Areces and from MEC grant BFM2000-1312.

\begin{table}
\caption[]{ Total energies (calculated with a double-$\zeta$ (DZ)
and a double-$\zeta$ polarized (DZP) basis set) of structures of the
Si(557)-Au surface differing in the positions of the Au rows.  
Structure numbers refer to the
labelling defined in Fig.\ref{fig1}-(b), and the energies, in eV per
Au atom, are relative to model 2.}
\begin{tabular}{c|ccccc}
 & \multicolumn{1}{c}{Edge} & 1& 2& 3 & 4 \\
 DZP & 0.55  & 0.30 & 0.00  & -0.07 & -0.03  \\
 DZ  & 0.53  & 0.30 & 0.00  &  0.01 & -0.07
\end{tabular}
\label{table1}
\end{table}


\begin{references}
%
\bibitem[*]{byline} Email address: sqbsapod@scsx03.sc.ehu.es
%
\bibitem{Tomonaga} 
S. Tomonaga, Prog. Theor. Phys. {\bf 5}, 544 (1950)
%
\bibitem{Luttinger}
J.M. Luttinger, J. Math. Phys. {\bf 4}, 1154 (1963) 
%
\bibitem{Haldane}
F.D.M. Haldane J. Phys. C {\bf 14}, 2585 (1981).
%
\bibitem{Fisher} 
M.P.A. Fisher and A. Glazman, {\it Mesoscopic
Electron Transport} (Kluwer Academic, Boston (1997)).
%
\bibitem{NT} M. Bockrath {\it et al.}, Nature {\bf 395}, 598 (1998)  
%
\bibitem{Hall} A. M. Chang, L. N. Pfeiffer and, K. W. West, 
Phys. Rev. Lett. {\bf 77},
2538 (1996); A. M. Chang {\it et al.}, {\it ibid.} {\bf 86},
143 (2000).
%
\bibitem{wires} O. M. Auslaender {\it et al.}, Phys. Rev. Lett.
{\bf 84}, 1764 (2000).
%
\bibitem{organic} A. Schwartz {\it et al.}, Phys. Rev. B {\bf 58},
1261 (1998)
%
\bibitem{PrBaCuO} K. Takenaka {\it et al.} Phys. Rev. Lett. {\bf 85}, 5428 (2000).
%
\bibitem{SrCuO} C. Kim {\it et al.} Phys. Rev. Lett. {\bf 77}, 4054 (1996)
%
\bibitem{LiMoO} J.D. Denlinger {\it et al.} Phys. Rev. Lett. {\bf 82}, 2540 (1999).
%
\bibitem{Xue} J. Xue {\it et al.} Phys. Rev. Lett. {\bf 83}, 1235 (1999).
%
\bibitem{NatureSegovia}
P. Segovia, D. Purdie, M. Hengsberger, and Y. Baer,
Nature {\bf 402}, 504 (1999). 
%
\bibitem{Natureswires}
H. Onishi, Y. Kondo, and K. Takayanagi,
Nature {\bf 395} 780 (1998);
A.I.Yanson {\it et al.}, 
Nature {\bf 395}, 783 (1998).
%
\bibitem{PRL} 
D. S\'anchez-Portal  {\it et al.},
Phys. Rev. Lett. {\bf 83}, 3884 (1999). 
%
\bibitem{privateHimpsel} 
R. Losio {\it et al.}, Phys. Rev. Lett. {\bf 86}, 4632 (2001).
%
\bibitem{SIESTA} 
D. S\'anchez-Portal, 
P.Ordej\'on, E. Artacho, and J. M. Soler,
Int. J. of Quantum Chem. {\bf 65}, 453 (1997);
E. Artacho {\it et al.},
Phys. Status Solidi B {\bf 215} 809 (1999).
%
\bibitem{Troullier-Martins} N. Troullier and J.L. Martins,
Phys. Rev. B {\bf 43}, 1993 (1991).
%
\bibitem{basis}
Basis orbitals were generated with an {\it energy shift}
of 200~meV. Double-$\zeta$
(DZ) basis include two radial shapes 
for the 3$s$ and 3$p$ orbitals of Si, 6$s$ and 5$d$ of
Au, and 1$s$ of H. The DZP basis has an additional $d$ 
shell for Si, and $p$ shells for Au and H.
%
\bibitem{Mahony} J.D. O\'{ }Mahony {\it et al.}, 
Phys. Rev. B {\bf 49}, 2527 (1994)
%
\bibitem{Shibata} M. Shibata, I. Sumita, and M. Nakajima,  
Phys. Rev. B {\bf 57}, 1626 (1998).
%
\bibitem{PRLHimpsel}
R. Losio, K. N. Altmann, and F. J. Himpsel,
Phys. Rev. Lett. {\bf 85}, 808 (2000).
%
\bibitem{Mulliken} R. S. Mulliken, J. Chem. Phys. {\bf 23},
1841 (1955)
%
\bibitem{footnoteRMM}
We have also explored the effects of adding
a chain of adatoms in model 2, partially saturating
the dangling bonds in rows 3 and 4, as suggested 
by some STM images\cite{Shibata}. This inclusion 
is slightly favorable, with an energy gain of 
$\sim$0.1~eV/Au, but besides the
doubling of the unit cell it
does not produce any significant change in the band
structure, the Fermi level still pinned by the edge state
band. 
In this case there is a very small ($\sim$0.1~\AA)
dimerization of the Au chain.
%


\end{references}
\end{document}